\documentstyle[preprint,aps]{revtex}
\begin{document}
\draft
\title{Universality Classes for Interface Growth with Quenched Disorder}

\author{L. A. N. Amaral, A.-L. Barab\'asi and H. E. Stanley}

\address{Center for Polymer Studies and Dept. of Physics,
 Boston University, Boston, MA 02215 USA}

\date{\today}

\maketitle

\begin{abstract}
We present numerical evidence that there are two distinct universality
classes characterizing driven interface roughening in the presence of
quenched disorder.  The evidence is based on the behavior of
$\lambda$, the coefficient of the nonlinear term in the growth
equation.  Specifically, for three of the models studied, $\lambda
\rightarrow \infty$ at the depinning transition, while for the two
other models, $\lambda \rightarrow 0$.
\end{abstract}


\narrowtext

	The motion of a nonequilibrium interface in a disordered
environment has attracted much attention. Fluid flow in a porous
medium is a typical experimental realization of these phenomena, but
applications range from wetting phenomena to the motion of flux lines
in the presence of disorder \cite{books}.  The origin of different
universality classes is well understood for growth in which thermal
(or time-dependent) noise dominates the roughening process.  However,
recently several independent studies
\cite{books,theory,Robbins,Tang,Buldyrev,Parisi,Leschhorn,Csahok} have
noted that {\it quenched\/} noise --- which is independent of time and
depends only on the position of the interface --- may change the
universality class, generating interfaces with anomalously large
roughness exponents.

        In the typical case, a $d$-dimensional interface characterized by
a height $h(x,t)$ moves in a $(d+1)$-dimensional disordered medium. The
randomness of the medium can be described by a quenched noise
$\eta(x,h)$.  In the presence of an external driving force $F$, the
simplest growth equation describing the zero-temperature dynamics of the
interface is \cite{theory}
\begin{equation}
\partial_t h = F + \nu \nabla^2 h + \eta(x,h).
\label{ew}
\end{equation}
The $\nu \nabla^2 h$ term mimics a surface tension and acts to {\it
smooth\/} the interface, while the quenched noise $\eta(x,h)$ works to
{\it roughen\/} the interface.  It is generally assumed that the
quenched noise has zero mean and is uncorrelated.

	An interface characterized by (\ref{ew}) moves with a finite
velocity $v_0$ if the driving force exceeds a critical value $F_c$,
while for $F< F_c$ it is pinned by the disorder. When $F \rightarrow
F_c$, one finds
\begin{equation}
v_0 \sim f^\theta,
\label{vel}
\end{equation}
where $f \equiv (F-F_c)/F_c$ is the reduced force and $\theta$ is the
velocity exponent.

	Recently, a number of analytical \cite{theory} and numerical
\cite{Robbins,Tang,Buldyrev,Parisi,Leschhorn} studies focused on
understanding the nature of the depinning transition and obtaining
accurate estimates for the critical exponents.  Renormalization group
(RG) studies \cite{theory} of Eq. (\ref{ew}) find a roughness exponent
$\alpha =(4-d) /3$, but a number of numerical models
\cite{Robbins,Tang,Buldyrev,Parisi,Leschhorn} revealed exponents whose
values can be quite different from the RG predictions.  Wetting fluid
invasion models gave $\alpha \simeq 0.8$ for $(1+1)$ dimension, and
investigation of the random field Ising model (RFIM) in $(2+1)$
dimension gave $\alpha \simeq 0.67$ \cite{Robbins}.  Solid-on-solid type
models gave $\alpha \simeq 0.63$ for $(1+1)$ dimension, and $\alpha
\simeq 0.48$ for $(2+1)$ dimension \cite{Tang,Buldyrev}, while a
discretized solid-on-solid model of Eq. (\ref{ew}) gave $\alpha
\simeq 1.25$ for $(1+1)$ dimension, and $\alpha \simeq 0.75$ for
$(2+1)$ dimension \cite{Leschhorn}.  A similar scattering is found for
the values of the other exponents characterizing the depinning
transition ($F=F_c$).

	Here we report simulations of five distinct models that have
been introduced to investigate the motion of an interface in the
presence of quenched disorder
\cite{Robbins,Tang,Buldyrev,Parisi,Leschhorn}.  Our findings suggest the
existence of two different universality classes.  One universality
class is described by the nonlinear growth equation
\cite{KPZ86}
\begin{equation}
\partial_t h = F + \nu \nabla^2 h + \lambda (\nabla h)^2 + \eta(x,h),
\label{kpz}
\end{equation}
and we find that for models in this universality class
\begin{equation}
\lambda \sim f^{-\phi}.
\label{phi}
\end{equation}
The second universality class is described, at the depinning transition,
by (\ref{ew}).  We propose that the existence of the two universality
classes is the origin of the systematic differences between the
exponents predicted by RG calculations and estimates from numerical
simulations.  We find that measuring $\lambda$ can give information
about the universality classes to which a given growth process belongs.

	To calculate $\lambda$, we follow Ref. \cite{Krug90a} and
impose a ``tilt'' of slope $m$ on the interface.  For a
$(1+1)$-dimensional system, we consider a lattice with $L$ columns,
and ``build in'' the tilt by implementing helicoidal boundary
conditions, $h(0,t)=h(L,t)-Lm$ and $h(L+1,t)=h(1,t)+Lm$.  For a
$(2+1)$-dimensional system, we use the same boundary conditions, so
the tilt occurs only in one direction.

	For a model described by Eq. (\ref{ew}), the local velocity $v$
is {\it independent\/} of the tilt.  However, if a nonlinear term is
present in addition to the linear term, then from (\ref{kpz}) follows
that \cite{Krug90a}
\begin{equation}
v = v_0 + \lambda m^2.
\label{vel1}
\end{equation}
Thus by varying the tilt $m$, we can test for the presence of nonlinear
terms in the growth equation and calculate the coefficient $\lambda$.

	First, we treat the model introduced in Ref. \cite{Tang},
for which it was shown, in $(1+1)$ dimension, that the interface at
$F_c$ is pinned by a directed percolation (DP) cluster
\cite{Tang,Buldyrev}, and that the critical dynamics are
controlled by a divergent correlation length parallel to the interface
$\xi_{\parallel} \sim f^{-\nu_{\parallel}}$ with $\nu_{\parallel}
\simeq 1.73$.  This model, referred to as ``DP-1'', excludes overhangs,
and gives rise to a self-affine interface at the depinning transition,
with a roughness exponent $\alpha \simeq 0.63$ \cite{explain0}.

	In the DP-1 model, in $(1+1)$ dimension, we start from a
horizontal interface at the bottom edge of a lattice of size $L$.  At
every site of the lattice we define a random, uncorrelated quenched
variable, the noise $\eta$, with magnitude in the range
$[0,1]$. During the time evolution of the interface, we choose one of
the $L$ columns at random.  If the difference in height to the lowest
neighbor is larger than $(+1)$, this lowest neighboring column grows
by one unit.  Otherwise, the chosen column grows one unit provided the
noise on the site above the interface is smaller than the driving
force $F$.  The unit time is defined to be $L$ growth attempts.

	We measure the velocity of the interface for different reduced
forces $f$ and different tilts $m$.  The results for $(1+1)$ dimension
are shown in Fig. \ref{parab}a.  For a fixed force $f$, we find that the
interface velocity changes with the tilt $m$, indicating the existence
of nonlinear terms.  Near the depinning transition ($f \rightarrow 0$),
the velocity curves become ``steeper'' and from (\ref{vel1}), we infer
that $\lambda$ must increase.

	To measure $\lambda$, we first attempt to fit a parabola to
the tilt-dependent velocities in the vicinity of zero tilt. The
calculations indicate that as we approach the depinning transition,
$\lambda$ diverges according to Eq. (\ref{phi}).  However, in the
vicinity of $F_c$, the velocity curves lose their parabolic shape for
large tilts (see Figs \ref{parab}a and \ref{cross}), indicating the
presence of other terms not included in (\ref{vel1}).

	We can understand the breakdown of (\ref{vel1}) for large $m$
using scaling arguments.  Substituting Eqs. (\ref{vel}) and
(\ref{phi}) into (\ref{vel1}), we find
\begin{equation}
v(m, f) \propto  f^\theta + a  f^{-\phi} m^2.
\label{velo}
\end{equation}
Equation (\ref{velo}) indicates that the velocity curves corresponding
to two different forces $f_1$ and $f_2$, with $f_1>f_2$, will
intersect at a tilt $m_{\times}$ (see Fig. \ref{cross}). For tilts
greater than $m_{\times}$, $v(m, f_1) < v(m, f_2)$, a clearly
unphysical prediction; the average velocity, for the same tilt, should
be larger for the larger force.  Thus the velocity cannot follow a
parabola for arbitrarily large $m$, and a crossover to a different
behavior than that of Eq. (\ref{velo}) must occur for values of the
tilt larger than $m_{\times}$.

	Letting $(f_1 - f_2) \rightarrow 0$, we find from (\ref{velo})
that the crossing point of the two corresponding parabolas scales as
\begin{equation}
m_{\times}^2 \sim f^{\theta + \phi}.
\label{range}
\end{equation}
Equations (\ref{velo}) and (\ref{range}) motivate the scaling form for
the velocities
\begin{equation}
v(m, f) \sim f^\theta g(m^2 / f^{\theta + \phi}),
\label{scal}
\end{equation}
where $g(x) \sim const.$ for $x \ll 1$, and $g(x) \sim
x^{\theta/(\theta + \phi)}$ for $x \gg 1$ \cite{explain00}.  Figure
\ref{figscal}a shows the data collapse we obtain using (\ref{scal}),
and the results of Fig. \ref{parab}a rescaled with exponents $\theta=
0.64 \pm 0.08$, $\phi= 0.64 \pm 0.08$ for $(1+1)$ dimension
\cite{explain1}

	The scaling behavior (\ref{scal}) is not limited to the DP-1
model in $(1+1)$ dimension, for $(2+1)$ dimension and for the models
introduced in Refs. \cite{Buldyrev,Parisi} we find a very similar
behavior. We refer to these models as ``DP-2'' \cite{Buldyrev} and
``Parisi'' \cite{Parisi}.  We simulated them in $(1+1)$ dimension, and
were able to rescale the velocities according to (\ref{scal}) using the
exponents presented in Table \ref{tab1}.

	Another model we studied was the RFIM, which allows for
overhangs; and for certain values of its parameters can be mapped to
percolation \cite{Robbins}.  In the RFIM, spins on a square lattice
interact through the Hamiltonian
\begin{equation}
{\cal H} \equiv - \sum_{\langle i,j\rangle} S_iS_j - \sum_i
[F+\eta(i,h)]S_i,
\label{rfim}
\end{equation}
where $S_i=\pm1$, $F$ now denotes the external magnetic field, and
$\eta$ is the time-independent local random field (i.e., quenched
noise) whose values are uniformly distributed in the interval
$[-\Delta,\Delta]$.  The strength of the quenched disorder is
characterized by the parameter $\Delta$.  At time zero, all spins are
``down''---except those in the first row, which are initially up.
During the time evolution of the system, we flip any down spin that is
``unstable,'' i.e., whenever the flip will lower the total energy of
the system.  The control parameter of the depinning transition is the
external magnetic field $F$; the unit time corresponds to flipping all
unstable spins \cite{explain2}.

	For dimension $(1+1)$, there are two morphologically-different
regimes, depending on the strength $\Delta$ of the disorder (i.e., of
the random fields).  For $\Delta > 1.0$, the interface is self-similar
(SS), while for $\Delta < 1.0$ it is faceted (FA).  For dimension
$(2+1)$, there is again a FA regime ($\Delta < 2.4$), a SS regime
($\Delta > 3.4$), and also a self-affine (SA) regime in between ($2.4 <
\Delta < 3.4$) \cite{Robbins}.  The SA regime, which exists only for
$(2+1)$ dimension, is the only regime of the RFIM for which either Eqs.
(\ref{ew}) or (\ref{kpz}) could apply.  In the SS regime, the interface
is not single-valued, while in the FA regime, lattice effects dominate the
growth.

	Our results show that for the FA regime, the RFIM behaves in a
similar fashion to the other three models, in that the coefficient of
the nonlinear term diverges at the depinning transition.  However,
although (\ref{scal}) is still valid for the SA and SS regimes, we
find a negative $\phi$, thus $\lambda \rightarrow 0$.  This behavior
can be understood, for the SS regime, by considering that near the
depinning transition, the morphology of the interface corresponds to
the hull of a percolation cluster, which has no well-defined
orientation \cite{Robbins}.  Thus a change in the boundary conditions
will not affect the growth process, and we cannot expect any
divergence of a possible nonlinear term when the magnetic field
approaches its critical value.  On the other hand, for large fields,
the effect of the quenched disorder diminishes, and we can observe an
average interface orientation.  For such values of field, we expect
the presence of nonlinear terms to be felt.  Although for the SA
regime the behavior of $\lambda$ is similar (see Figs. \ref{parab}b
and \ref{figscal}b), the reasons so far cannot be understood.

	These results lead us to conclude that in the SA regime the RFIM
belongs to the universality class of Eq.  (\ref{ew}).  This conclusion
is further supported by the agreement between the numerically determined
exponents, $\alpha \simeq 0.67$ and $\theta \simeq 0.60$ for $(2+1)$
dimension, and the RG predictions, $\alpha = \theta = 2/3$
\cite{theory}.

	Finally we studied the discretized solid-on-solid version of
Eq.  (\ref{ew}), referred to as ``SOS-1'' \cite{Leschhorn}, and find
that for any reduced force, $\lambda = 0$.

	The results of Table \ref{tab1} show, for $(1+1)$ dimension, a
clear separation into two groups in the values of the critical
exponents for the five models studied.  In the following we argue that
this separation reflects the existence of two distinct universality
classes, described by the two continuum growth equations, (\ref{ew})
and (\ref{kpz}).  For the SOS-1 model and for the RFIM, in the SA regime,
we find that $\lambda$ either is zero or goes to zero at the depinning
transition.  Thus the scaling behavior of these models should be
correctly described by (\ref{ew}).  For the DP-1, DP-2 and Parisi models
we observe a divergent $\lambda$, indicating that nonlinearities are
relevant near the depinning transition.  Thus to properly describe the
scaling properties of these models it is necessary to study
(\ref{kpz}), since (\ref{ew}) does not includes the nonlinear term
$\lambda (\nabla h)^2$.  Further evidence of the existence of the two
universality classes is given by the values of roughness exponents.
The models for which $\lambda$ diverges at the depinning transition
\cite{Tang,Buldyrev,Parisi}, predict $\alpha \simeq 0.63$, in
agreement with the mapping to DP.  On the other hand, models in the
universality class of Eq. (\ref{ew}) \cite{Robbins,Leschhorn}, gave
roughness exponents typically larger, in better agreement with the RG
predictions \cite{theory}.  Finally, we propose the study of the
behavior of $\lambda$ at the depinning transition as a general method
for identifying the universality class of a given growth process in
disordered media.

	This work is part of the Ph.D. thesis of L. A. N. Amaral, to
be submitted to Boston University.  We thank R. Cuerno, S. Havlin, H.
Makse, O. Narayan, T.  Nattermann and especially S. V.  Buldyrev for
useful suggestions.  L. A. N.  Amaral acknowledges a scholarship from
Junta Nacional de Investiga\c c\~ao Cient\'{\i}fica e Tecnol\'ogica.
The Center for Polymer Studies is supported by the National Science
Foundation.

\begin{figure}
\caption{Dependence on the tilt $m$ of the average velocity, (a) in
the DP-1 model and (b) in the RFIM.  Data for different forces $f$ are
indicated by different symbols, ranging from $0.016$ (bottom curve) to
$0.350$ (top curve) for the DP-1 model, and from $0.014$ (bottom
curve) to $0.143$ (top curve) for the RFIM. In (a) we show velocities,
for the DP-1 model, for a system of size $512$ in $(1+1)$ dimension.
In (b) are plotted the velocities for the RFIM in the SA regime
($\Delta = 3$), for a $(2+1)$-dimensional system of size $40 \times
40$.}
\label{parab}
\end{figure}

\begin{figure}
\caption{Here we exemplify the ``noncrossing'' effect on the velocity
parabolas. We show a perfect parabolic behavior for two different
forces, $f_1>f_2$ (dashed lines) as predicted by Eq. (5).  Also shown
is the ``curving back'' of the velocity curve for the smaller force
$f_2$ (solid line) in order not to cross the velocity curve for
$f_1$.}
\label{cross}
\end{figure}

\begin{figure}
\caption{Data collapse according to (8), using the same symbols for
the velocities shown in Fig.  1.  In (a) we present the rescaled
results for the DP-1 model in $(1+1)$ dimension and in (b) the
rescaling of the $(2+1)$-dimensional results for the RFIM, in the SA
regime ($\Delta = 3.0$).}
\label{figscal}
\end{figure}

\begin{table}
\caption{Exponents for the five studied models (see definitions in
the text).  A negative value of $\phi$ means that $\lambda
\rightarrow 0$ when $f \rightarrow 0$.  We argue in the text that the
models above the horizontal line (DP-1, DP-2, and Parisi) belong to
the universality class of Eq. (3) and can be mapped, in $(1+1)$
dimension, to DP.  The models below the line belong to the
universality class of Eq. (1).}
\begin{tabular}{llcccc}
\tableline
\multicolumn{2}{l}{Model} & \multicolumn{2}{c}{$(1+1)$ dimension} &
  \multicolumn{2}{c}{$(2+1)$ dimension} \\ & & $\theta$ & $\phi$ &
$\theta$ & $\phi$ \\
\tableline
\tableline
DP-1 	&	& $0.64\pm0.08$	& $0.64\pm0.08$	& $0.80\pm0.12$	&
$0.30\pm0.12$	\\
DP-2 	&	& $0.59\pm0.12$	& $0.55\pm0.12$ &	& 	\\
Parisi	&	& $0.70\pm0.12$ & $0.65\pm0.12$	& 	&	\\
\tableline
RFIM	& SA 	& ---	& ---	& $0.60\pm0.11$ & $-0.70\pm0.11$
\\
	& SS	& $0.31\pm0.08$ & $-0.65\pm0.13$&	& 	\\
SOS-1	&	& $0.26\pm0.07$ & ---		& 	&	\\
\tableline
\end{tabular}
\label{tab1}
\end{table}


\begin{thebibliography}{10}
\bibitem{explain0}
The mapping to DP allows us to calculate the roughness exponent
$\alpha$ at the depinning transition [4,5].  Near the transition the
width of the interface $w$ is given by the perpendicular correlation
length $\xi_{\perp}$, and the parallel correlation length
$\xi_{\parallel}$ is of the order of the system size.  So $w \sim
L^{\nu_{\perp}/\nu_{\parallel}}$, giving for the roughness exponent
$\alpha = \nu_{\perp}/\nu_{\parallel} \simeq 0.63$.
\bibitem{explain00}
The consistency of the argument developed in the text can be verified
by expanding (8) to first-order, whereupon we recover (6).
\bibitem{explain1}
The value of the velocity exponent can also be obtained by considering
the mapping to DP [4].  The average velocity close to the depinning
transition scales as $\xi_{\perp} / \xi_{\parallel}$, giving $\theta =
\nu_{\parallel} - \nu_{\perp} \simeq 0.64$.
\end{thebibliography}
\end{document}